\documentclass[journal=jacsat,manuscript=article]{achemso}  
\usepackage{amsfonts}
\usepackage{amssymb}
\usepackage{graphicx}
\usepackage{amsmath}
\usepackage{tikz}
\usepackage{adjustbox}
\usepackage[normalem]{ulem}
\usetikzlibrary{shapes.geometric, arrows}
\usepackage{indentfirst}

\usepackage[english]{babel}
\usepackage{color}
\usepackage[version=3]{mhchem} 
\usepackage{natbib}
\usepackage{url}
\usepackage[colorlinks,citecolor=black,urlcolor=blue,linkcolor=black]{hyperref} 
\usetikzlibrary{arrows,shapes,positioning,shadows,trees}

\newcommand{\olr}[1]{{\color{red}{}}}

\newcommand{\osvp}[1]{{\color{blue}{}}}

\author{Sebastian V. Pios}
\affiliation{Zhejiang Laboratory, Hangzhou 311100, China}
\author{Maxim F. Gelin}
\affiliation{School of Science, Hangzhou Dianzi University, Hangzhou 310018, China}
\author{Wolfgang Domcke}
\affiliation{Department of Chemistry, Technical University of Munich, D-85747 Garching, Germany}
\author{Lipeng Chen}
\affiliation{Zhejiang Laboratory, Hangzhou 311100, China}
\email{chenlp@zhejianglab.com}

\title{Imaging the Photochemistry of the Hydrogen-Bonded Heptazine-Water Complex with Femtosecond Time-Resolved Spectroscopy: A Computational Study}

\begin{document}


\begin{abstract}
Graphitic carbon nitride (\textit{g}-CN) has attracted vast interest as a promising inexpensive metal-free photocatalyst for water splitting with solar photons. 
The heptazine (Hz) molecule is the building block of graphitic carbon nitride. 
The photochemistry of the Hz molecule and derivatives thereof in protic environments has been the subject of several recent experimental and computational studies. 
In the present work, the hydrogen-bonded Hz$\cdots$H\textsubscript{2}O complex was adopted as a model system for the exploration of photoinduced electron and proton transfer processes in this complex with quasi-classical nonadiabatic trajectory simulations, using the \textit{ab initio} ADC(2) electronic-structure method and a computationally efficient surface-hopping algorithm. 
The population of the optically excited bright $^{1}\pi\pi^{*}$ state of the Hz chromophore relaxes through three $^{1}n\pi^{*}$ states and a low-lying charge-transfer state, which drives proton transfer from H\textsubscript{2}O to Hz, to the long-lived optically dark S\textsubscript{1}($\pi\pi^{*}$) state of Hz. 
The imaging of this ultrafast and complex dynamics with femtosecond time-resolved transient absorption (TA) pump-probe (PP) spectroscopy and two-dimensional (2D) electronic spectroscopy (ES) was computationally explored in the framework of the quasi-classical doorway-window approximation. 
By comparison of the spectra of the Hz$\cdots$H\textsubscript{2}O complex with those of the free Hz molecule, the effects of the hydrogen bond on the ultrafast internal conversion dynamics can be identified in the spectroscopic signals. 
Albeit the TA PP and 2D ES spectroscopies are primarily sensitive to electronic excited-state dynamics and less so to proton transfer dynamics, they nevertheless can provide mechanistic insights which can contribute to the acceleration of the optimization of photocatalysts for water splitting.
\end{abstract}

\section{INTRODUCTION}

Graphitic carbon nitride (\textit{g}-C\textsubscript{3}N\textsubscript{4}) and related materials have received extensive attention in the past decade as metal-free photocatalysts for hydrogen evolution from water with sacrificial reagents.\cite{wang2009metal,wang_angewandte_2012,ong_chemrev_2016,WEN2017} 
These amorphous or partially crystalline materials have attractive properties with respect to UV/vis light absorption, exciton dissociation and charge carrier migration to interfaces, where water molecules are oxidized by holes and protons are neutralized by electrons to yield gaseous H\textsubscript{2} and O\textsubscript{2}.\cite{ong_chemrev_2016,WEN2017}

The molecular building block of \textit{g}-C\textsubscript{3}N\textsubscript{4} is heptazine (heptaazaphenalene), denoted as Hz in what follows. 
While Hz itself is chemically and photochemically unstable due to rapid hydrolysis in the presence of traces of water, a derivative of Hz, 2,5,8-tris(4-methoxyphenyl)-heptaazaphenalene or trianisole-heptazine (TAHz), was shown to be chemically highly stable under UV irradiation.\cite{rabe_tahz_2018} 
Schlenker and coworkers studied the spectroscopy and photochemical kinetics of TAHz in toluene and in liquid water and demonstrated the liberation of OH radicals from water under UV/vis irradiation. 
The elementary reaction mechanisms of TAHz in protic solvents were further characterized by studying the spectroscopy and photochemistry of hydrogen-bonded complexes of TAHz with substituted phenols with tuneable oxidation potentials.\cite{rabe_jpcc_2019} 
It was demonstrated that the barrier for the excited-state proton-coupled electron-transfer (PCET) reaction (from phenol to TAHz) could be tuned by chemical substitutions of phenol while the PCET reaction could be actively controlled by a picosecond push pulse in a pump-push-probe experiment.\cite{corp_jpcc_2020} 
The interpretation of the data was supported by \textit{ab initio} electronic-structure calculations of excited-state reaction paths and reaction barriers.\cite{schlenker_overview2021,domcke_water_ox_2022} 


The elementary photophysical and photochemical processes involved in hydrogen evolution with polymeric carbon nitrides are not well understood due to the complexity of these materials and their poorly constrained structure at the atomic level. 
Measurements of nonlinear spectroscopic signals with femtosecond time resolution are challenging due to the high optical density which prevents detection of signals in transmission, making experiments easier to realize for dilute solutions of molecular Hz chromophores rather than for bulk material. 
With currently available technology, it may be possible to characterize the kinetics of the photoinduced PCET reaction and competing relaxation processes in hydrogen-bonded complexes of Hz derivatives with water molecules.

Basic features of the PCET reaction in the Hz$\cdots$H\textsubscript{2}O complex (the dots denote a hydrogen bond) were recently explored with \textit{ab initio} on-the-fly nonadiabatic trajectory simulations using the ADC(2) electronic-structure method.\cite{xiang_heptazine2021}
In the present work, this previous study is extended by performing simulations of femtosecond time-resolved transient absorption (TA) pump-probe (PP) and two-dimensional (2D) electronic spectra (ES) for the Hz$\cdots$H\textsubscript{2}O complex using the quasi-classical implementation of the doorway-window (DW) approximation.\cite{OTFDW1,Xiang2D} 

In TA PP spectroscopy, two short laser pulses, a pump and a probe pulse, are used to excite the system and then to detect the transmission of the probe pulse through the sample as a function of the time delay \textit{T} (see Figure~S9a in the Supporting Information (SI) for a schematic picture).
It is the most commonly applied method of time- and frequency-resolved nonlinear spectroscopy.\cite{khundkar1990ultrafast,martin1992femtosecond,buckup2014multidimensional}

2D ES is a variant of third-order heterodyne-detected four-wave-mixing (FWM) spectroscopy.
The pulse sequence is schematically shown in Figure~S9b. 
The FWM signal $S^{(\sigma)}(\tau,T,\tau_{t})$ can be recorded in the rephasing ($\sigma=R$, $\mathbf{k_{S}}=\mathbf{k_{1}}-\mathbf{k_{2}}+\mathbf{k_{3}}$) or non-rephasing ($\sigma=NR$, $\mathbf{k_{S}}=-\mathbf{k_{1}}+\mathbf{k_{2}}+\mathbf{k_{3}}$) phase-matching directions as a function of the delay time between the first two pulses (coherence time $\tau$), the delay time between the second and the third pulse (population time $T$), and the time interval between the last two pulses (detection time $\tau_{t}$). 
Fourier transform of the FWM signal with respect to $\tau$ and $\tau_{\mathrm{t}}$ yields the 2D signal as function of the excitation frequency $\omega_{\tau}$ and detection frequency $\omega_{\mathrm{t}}$.
The 2D spectrum $S^{(\sigma)}(\omega_{\tau}, T, \omega_t)$ depicts the evolution of correlation between the initial ($\omega_{\tau}$) and final ($\omega_{t}$) state coherences as a function of the population time \textit{T}. 
2D ES yields signals with higher information content than TA PP spectroscopy since the excitation and emission frequencies are separately resolved.

Both TA PP spectroscopy and 2D ES have the feature of consisting of three contributions, the ground-state bleach (GSB), the stimulated emission (SE) and the excited state absorption (ESA) contributions.\cite{MukamelBook}
Unfortunately, GSB, SE and ESA spectra often overlap in frequency space, leading to the partial cancellation of the spectral information. 
While techniques such as polarization-sensitive detection \cite{Hochstrasser01,Gelin13a,Zanni22} or beating maps \cite{Leonas14,EET1,EET3} can partially alleviate the problem, they do not provide a universal solution. 

The DW formalism was first introduced by Mukamel and coworkers in 1989, \cite{Yan89} providing an intuitive way of interpreting TA PP and related four-wave-mixing signals.
In the DW picture, the pump pulse generates the so-called doorway operator in the electronic ground state and in excited states, followed by the time evolution of the doorway operator with the field-free Liouvillian.
The window operator is generated by the probe pulse.
The contraction of the time-evolved doorway operator with the window operator yields the signal. 

The quasi-classical approximation to the quantum DW representation was proposed by Fried and Mukamel\cite{fried1990} and was recently extended and recast in a form that is suitable for implementation with \textit{ab initio} on-the-fly trajectory methods.\cite{OTFDW1}
For a more detailed account of the quasi-classical DW approximation,  the reader is referred to Section S5 in the SI.

Apart from the demonstration of the feasibility of such first-principles simulations of nonlinear spectroscopic signals for a relatively large molecular system, the present work provides insight to which extent the photoinduced ultrafast electronic and nuclear dynamics of a chromophore-water complex can be visualized with femtosecond nonlinear spectroscopy.

\section{RESULTS}

\subsection{Electronic Structure, UV/Vis Absorption Spectrum and Nonadiabatic Excited-State Dynamics of Hz$\cdots$H\textsubscript{2}O}
\begin{figure}
    \centering
    \includegraphics[width=0.8\textwidth]{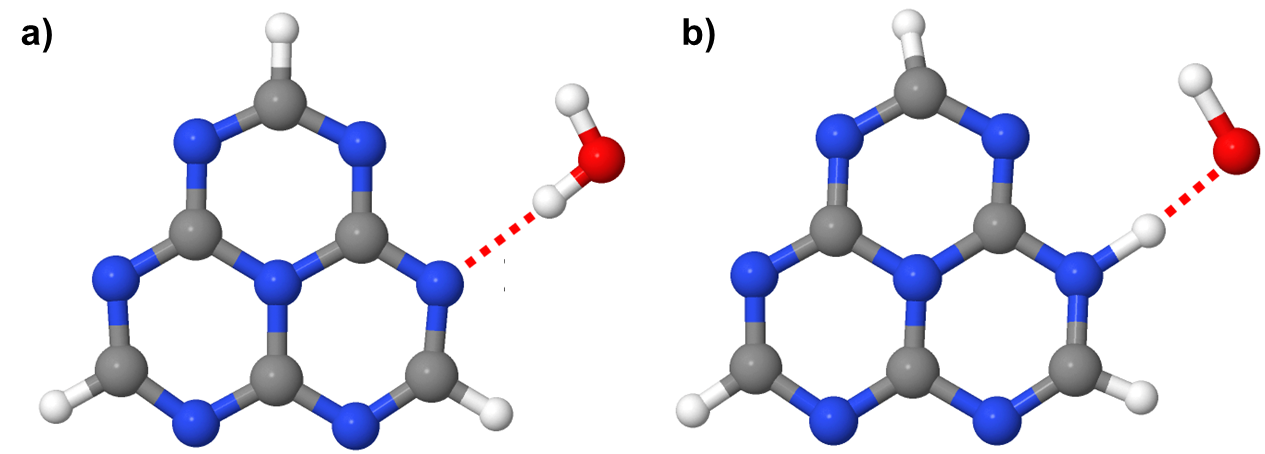}
    \caption{Ground-state equilibrium geometry of the hydrogen-bonded Hz$\cdots$H\textsubscript{2}O complex at the MP2 level (a). The hydrogen bond is indicated by the red dotted line. Representative geometry of the product of a successful PCET reaction (b).}
    \label{fig:ha_h2o_geo}
\end{figure}

The electronic structure calculations in the present work utilized the second-order M\o ller-Plesset  (MP2) method\cite{mp2_paper} and the algebraic diagrammatic construction scheme of second order (ADC(2)) for the polarization propagator\cite{adc2_paper1,adc2_paper2} (for computational details, see Section S1.1 in the SI).
The ground-state equilibrium geometry of the Hz$\cdots$H\textsubscript{2}O complex at the MP2 level is shown in Figure~\ref{fig:ha_h2o_geo}a. 
The length of the hydrogen bond (indicated by the dotted line) between Hz and the water molecule is 2.01~\AA. 
All atoms essentially lie in a plane with only tiny out-of-plane distortions.
Figure~\ref{fig:ha_h2o_geo}b shows a representative geometry of the HzH$\cdots$OH reaction product of a successful PCET reaction.
This geometry is for illustration only and has not been relaxed with respect to all degrees of freedom.

Table \ref{tab:vert_ex_energies} lists the seven lowest vertical excitation energies, oscillator strengths and dipole moments of the free Hz molecule and the Hz$\cdots$H\textsubscript{2}O complex.
In the isolated Hz molecule, the S\textsubscript{3} and S\textsubscript{4} as well as the S\textsubscript{5} and S\textsubscript{6} states have equal energies due to D\textsubscript{3h} symmetry. 
In the present calculations, these degeneracies are very slightly lifted due to the partial augmentation of the basis set which marginally breaks D\textsubscript{3h} symmetry (see Section S1.1 of the SI).
In Hz$\cdots$H\textsubscript{2}O, the degeneracy is additionally lifted by the hydrogen bond of Hz with the water molecule. 
The nondegenerate S\textsubscript{1} state of Hz arises from the excitation of an electron from the highest occupied orbital (HOMO) to the lowest unoccupied orbital (LUMO). 
The S\textsubscript{1}-S\textsubscript{0} transition is dipole forbidden in D\textsubscript{3h} symmetry. 

\begin{table}
\caption{Vertical excitation energies (in eV) with oscillator strengths (in parentheses) and permanent dipole moments (in brackets, given in Debye) of the lowest seven excited singlet states in the Hz molecule and the Hz$\cdots$H\textsubscript{2}O complex at ADC(2) level. The energy of the $\pi\pi^{*}$ charge-transfer state in Hz$\cdots$H\textsubscript{2}O is also listed.}
\begin{tabular}{c|c|c}\label{tab:vert_ex_energies}
                   & Hz                       & Hz$\cdots$H\textsubscript{2}O              \\ \hline
S\textsubscript{1} ($\pi\pi^*$) & 2.58 (0.000) {[}0.161{]} & 2.60 (0.000) {[}3.127{]}      \\
S\textsubscript{2} ($n\pi^*$)   & 3.75 (0.000) {[}0.427{]} & 3.76 (0.000) {[}6.545{]}      \\
S\textsubscript{3} ($n\pi^*$)   & 3.84 (0.000) {[}2.918{]} & 3.82 (0.000) {[}4.509{]}      \\
S\textsubscript{4} ($n\pi^*$)   & 3.87 (0.000) {[}2.705{]} & 3.95 (0.000) {[}1.234{]}      \\
S\textsubscript{5} ($\pi\pi^*$) & 4.39 (0.272) {[}1.033{]} & 4.38 (0.276) {[}3.812{]}      \\
S\textsubscript{6} ($\pi\pi^*$) & 4.41 (0.263) {[}0.741{]} & 4.42 (0.264) {[}2.850{]}      \\
S\textsubscript{7} ($n\pi^*$)   & 4.91 (0.000) {[}1.249{]} & 4.87 (0.002) {[}8.398{]}      \\
CT state           ($\pi\pi^*$) & -                        & 5.87 (0.017) {[}16.67{]}                       
\end{tabular}
\end{table}

\begin{figure}
    \centering
    \includegraphics[width=0.3\textwidth]{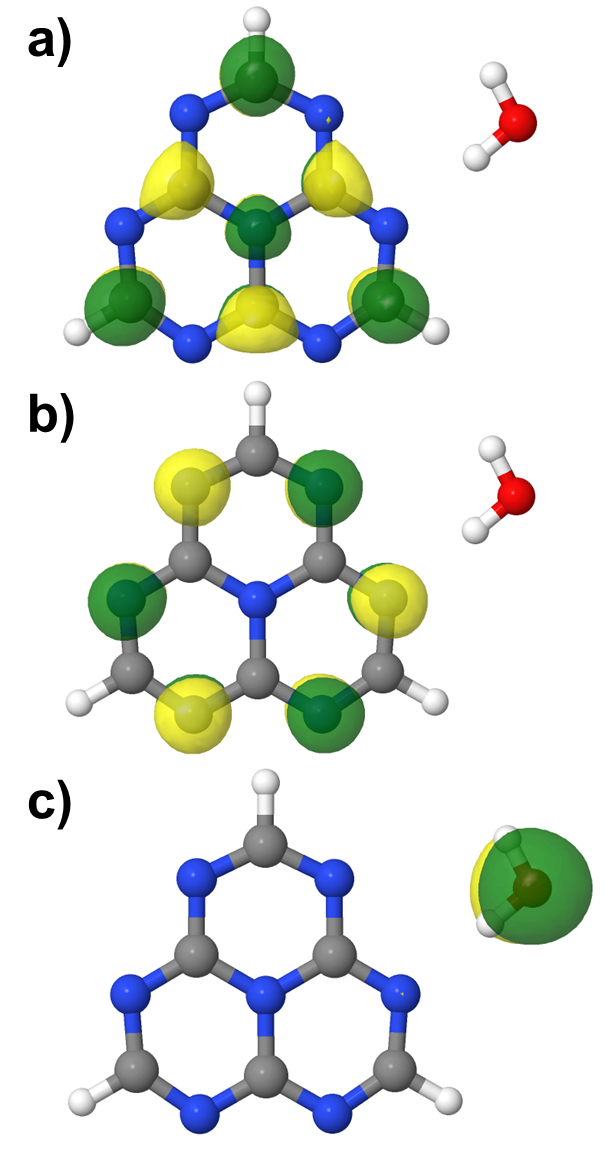}
    \caption{Selected molecular orbitals in the Hz$\cdots$H\textsubscript{2}O complex at the ground-state equilibrium geometry. The LUMO (a), HOMO (b) and the reactive \textit{p}\textsubscript{z}-orbital on the O atom of the water molecule (c) are shown.}
    \label{fig:hz_h2o_orbitals}
\end{figure}

The LUMO and HOMO of the Hz$\cdots$H\textsubscript{2}O complex are depicted in panels (a) and (b) of Figure~\ref{fig:hz_h2o_orbitals}, respectively. 
They are completely localized on Hz and are essentially identical with those of the isolated Hz molecule. 
The charge distribution of the highest nonbonding orbital of Hz, on the contrary, is significantly affected by the hydrogen bond (not shown). 
The next higher lying states (S\textsubscript{2}, S\textsubscript{3}, S\textsubscript{4}) are dark $n\pi^{*}$ states. 
As has been pointed out previously, there exist charge-transfer (CT) states in the Hz$\cdots$H\textsubscript{2}O complex which correspond to the transfer of an electron from \textit{n} (\textit{p}\textsubscript{x}, \textit{p}\textsubscript{y}) or $\pi$ (\textit{p}\textsubscript{z}) orbitals of the O-atom of H\textsubscript{2}O to the LUMO of Hz.\cite{ehrmaier_hz_2017}
The \textit{p}\textsubscript{z} orbital of H\textsubscript{2}O, which becomes the hole orbital in the CT state of $\pi\pi^{*}$ character, is depicted in Figure~\ref{fig:hz_h2o_orbitals}c. 
For a more complete in-depth discussion of the orbitals involved in the lowest excited states in Hz$\cdots$H\textsubscript{2}O the reader is referred to reference \cite{ehrmaier_hz_2017}.

The vertical excitation energies, oscillator strengths and dipole moments of the lowest 27 and 30 excited electronic states of Hz and Hz$\cdots$H\textsubscript{2}O, respectively, are listed in Table~S1 of the SI. 
Notably, the presence of the water molecule adds three additional singlet states within the first 30 excited states, two CT states from the water molecule to the Hz chromophore (S\textsubscript{15} and S\textsubscript{28}) and one CT state from Hz to the water molecule (S\textsubscript{26}).
The CT states stand out by their large dipole moments. 
There are also several $n\pi^{*}$ states of significant CT character, e.g. S\textsubscript{7} and S\textsubscript{10}.
The states with CT from the water molecule to Hz are of $\pi\pi^{*}$ or $n\pi^{*}$ character, while the state S\textsubscript{26} of opposite CT direction is of $\pi\sigma^{*}$ character (the $\sigma^{*}$ orbital is located on the water molecule). 
The vertical excitation energies of the locally excited states on Hz are only marginally affected by the presence of the water molecule.
The influence of the water molecule becomes noticeable, however, when the dipole moments of the electronic states are scrutinized.
Investigation of the \textit{x}-, \textit{y}- and \textit{z}-components of the dipole moments reveals that locally excited states with an increased dipole moment relative to the electronic ground state exhibit a partial transfer of electron density towards the water molecule, whereas a decreased dipole moment (e.g., in the state S\textsubscript{4}) indicates a transfer of electron density away from the water molecule.

%

\begin{figure}
    \centering
    \includegraphics[width=0.6\textwidth]{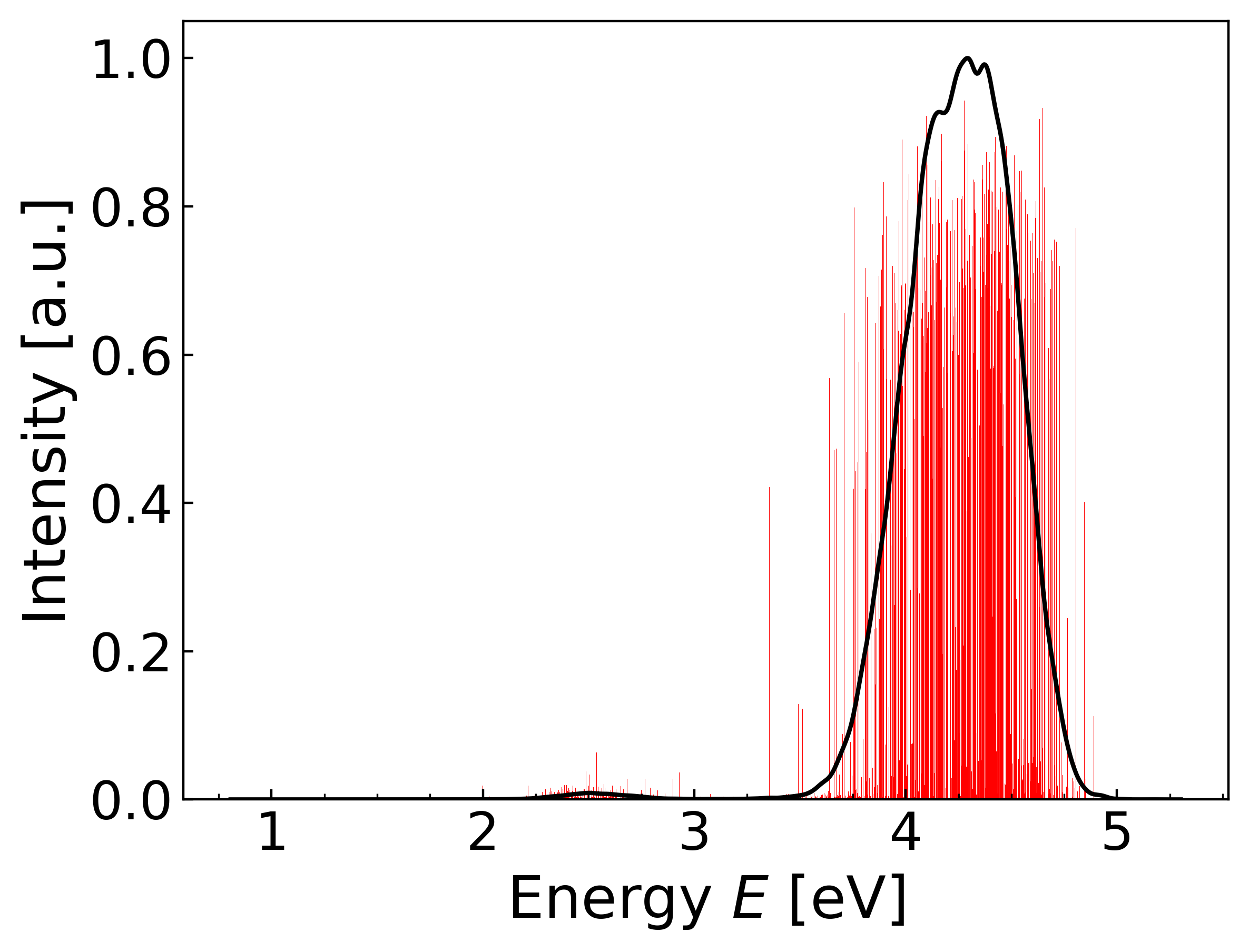}
    \caption{Absorption spectrum of the Hz$\cdots$H\textsubscript{2}O complex simulated with 5000 nuclear phase-space points selected from the ground-state Wigner distribution (black). The stick spectrum (red) was convoluted with a Gaussian of 0.05~eV FWHM.}
    \label{fig:hz_h2o_uvvis}
\end{figure}

The simulated UV/vis absorption spectrum of the Hz$\cdots$H\textsubscript{2}O complex is shown in Figure~\ref{fig:hz_h2o_uvvis}. 
The stick spectrum (red) represents the individual transition energies and intensities of the nuclear ensemble. 
The broad and intense peak centered at 4.29~eV represents the absorption of the bright S\textsubscript{5}/S\textsubscript{6} state. 
The very weak peak centered near 2.50~eV arises from the nominally forbidden S\textsubscript{1}-S\textsubscript{0} transition which becomes weakly allowed by vibronic intensity borrowing from the bright $^{1}\pi\pi^{*}$ state (Herzberg-Teller effect). 
The corresponding absorption spectrum of the isolated Hz molecule is displayed in Fig~S2. 
Apart from a blue-shift of the intense peak by 0.08~eV, it is identical with the spectrum in Figure~\ref{fig:hz_h2o_uvvis}.

%

\begin{figure}
    \centering
    \includegraphics[width=0.7\textwidth]{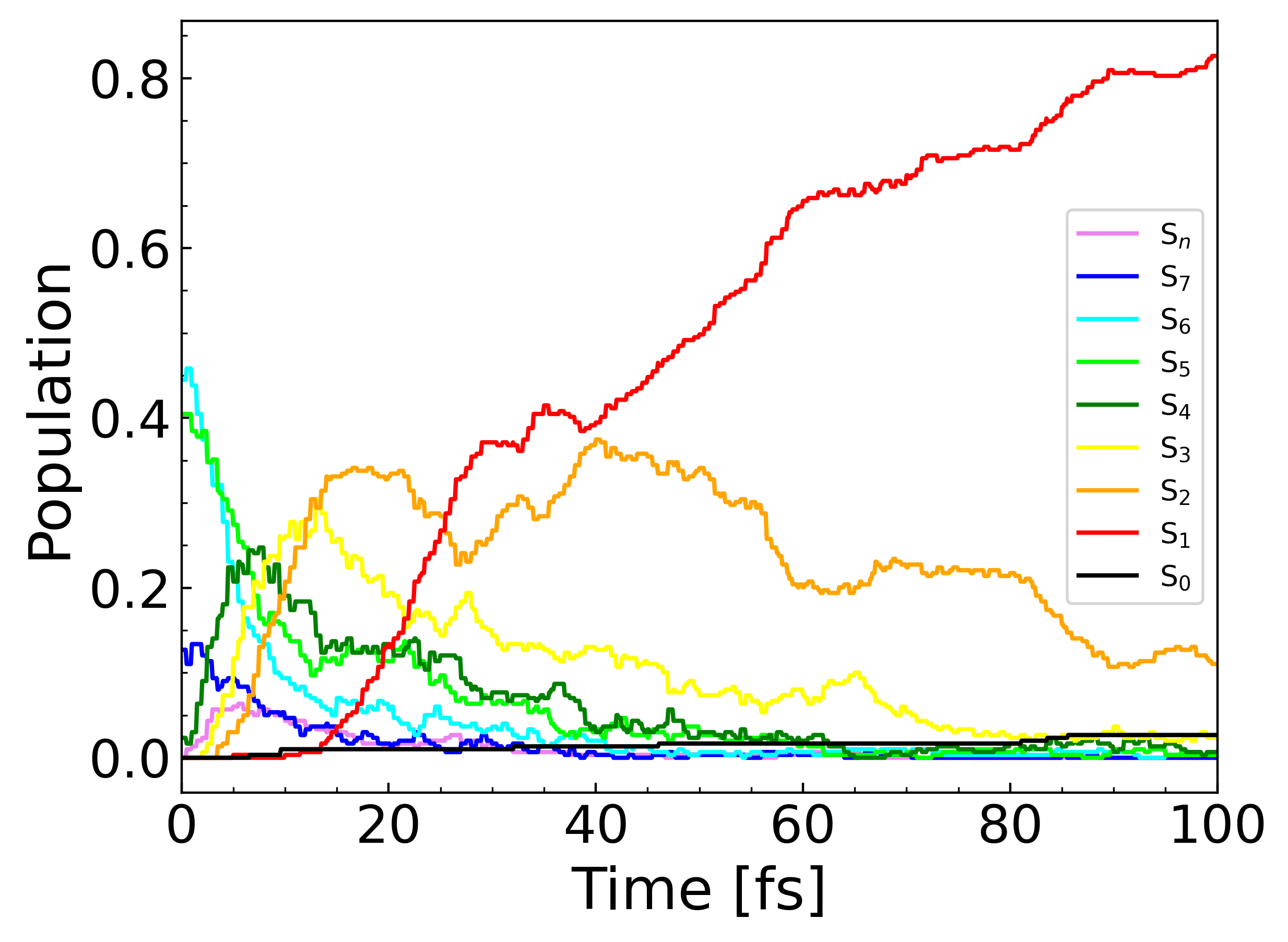}
    \caption{Population probabilities of the adiabatic singlet states in the Hz$\cdots$H\textsubscript{2}O complex up to 100~fs. See legend for the color code. The population of the electronic states S\textsubscript{\textit{n}} with \textit{n} $>$ 7 are collectively displayed and labeled S\textsubscript{\textit{n}}.}
    \label{fig:hz_h2o_adiabatic_pop}
\end{figure}

Figure~\ref{fig:hz_h2o_adiabatic_pop} depicts the time-dependent populations of the adiabatic electronic states of the Hz$\cdots$H\textsubscript{2}O complex for the first 100~femtoseconds after instantaneous photoexcitation. 
For clarity, the minor population probabilities of the states S\textsubscript{\textit{n}} with \textit{n} $>$ 7 are collectively shown as “S\textsubscript{\textit{n}}” in the figure.
These results are similar to previous work\cite{xiang_heptazine2021} and will be discussed only briefly. 
The initial conditions of the trajectories are selected from a 0.1~eV wide energy window at the maximum of the absorption spectrum. 
The initial photoexcitation populates mainly the two bright \textsuperscript{1}$\pi\pi^{*}$ states S\textsubscript{5}, S\textsubscript{6} and to a lesser extent the \textsuperscript{1}$n\pi^{*}$ state S\textsubscript{7}. 
The populations of S\textsubscript{5}, S\textsubscript{6}, S\textsubscript{7} decay on a time scale of about 10~fs, while the populations of the \textsuperscript{1}$n\pi^{*}$ states S\textsubscript{2} and S\textsubscript{3} rise on a similar time scale. 
The populations of S\textsubscript{2} and S\textsubscript{3} fluctuate and decay on time scales less than 100~fs. 
Beginning at about 20~fs, the population of the S\textsubscript{1} state of $\pi\pi^{*}$ character rises continuously. 
At the end of the propagation (100~fs), more than 80\% of the electronic population is in the S\textsubscript{1} state. 
It should be noted that the population of the electronic ground state (black) rises slowly and attains a value of about 3\% at 100 fs.

The adiabatic population probabilities of the singlet states of the isolated Hz molecule are displayed in Figure~S3. 
The differences in the populations of the electronic states S\textsubscript{4}, S\textsubscript{5} at time zero compared to those in Figure~\ref{fig:hz_h2o_adiabatic_pop} are due to the near degeneracy of these states and slightly different energy windows and are not of relevance. 
Overall, the time evolutions of the electronic population probabilities of Hz and Hz$\cdots$H\textsubscript{2}O are very similar. 
The water molecule does not absorb in this energy range and it seems that the perturbation of Hz by the hydrogen bond with the water molecule is too weak to affect the ultrafast electronic relaxation dynamics in the Hz chromophore. 
Notably, however, the S\textsubscript{0} state is not populated at all in isolated Hz, in contrast to the Hz$\cdots$H\textsubscript{2}O complex. 
While this effect may appear minor within the 100~fs time window of the present simulation, it will become significant on picosecond and nanosecond time scales. 
Indeed, it has been experimentally observed that the fluorescence lifetime of the S\textsubscript{1} state of TAHz decreases from 287~ns in toluene to 10~ns in water,\cite{rabe_tahz_2018} which is in qualitative agreement with the present results.

\subsection{Transient-Absorption Pump-Probe Spectrum}

For a detailed derivation of the doorway-window simulation protocol for TA PP spectra, see Sections S1.3 and S5 of the SI.  In all TA PP  calculations, the durations of the pump and probe pulses are fixed at $\tau_{\mathrm{pu}}$ = $\tau_{\mathrm{pr}}$ = 5~fs. 
Such short pulses are available in the UV/Vis domain in many spectroscopic labs worldwide.

\begin{figure}
    \centering
    \includegraphics[width=0.75\textwidth]{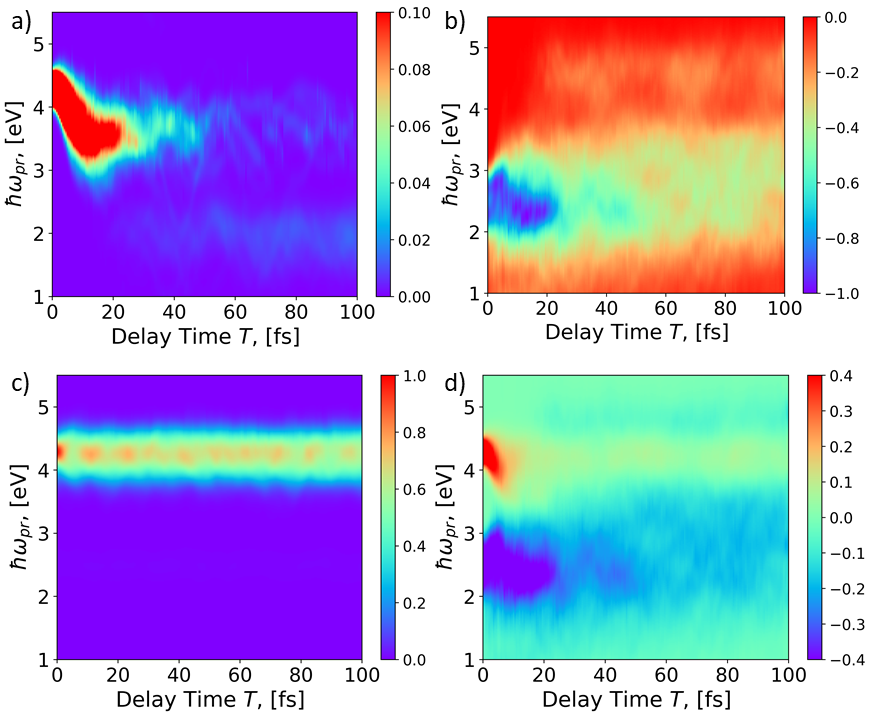}
    \caption{SE (a), ESA (b) and GSB (c) contributions and the total integral signal (d) as a function of delay time \textit{T} and probe energy $\hbar\omega_{\mathrm{pr}}$ for the Hz$\cdots$H\textsubscript{2}O complex. The pump pulse is in resonance with the maximum of the absorption spectrum ($\hbar\omega_{\mathrm{pu}}$ = 4.29~eV). The pulse widths are $\tau_{\mathrm{pu}}$ = $\tau_{\mathrm{pr}}$ = 5~fs. In the total signal (d), the GSB contribution has been scaled down by a factor of ten to improve the visibility of the other two contributions to the signal.}
    \label{fig:hz_h2o_integral}
\end{figure}

As mentioned above,\cite{MukamelBook} the integral TA PP signal consists of three contributions: ground-state bleach (GSB), stimulated emission (SE) and excited-state absorption (ESA). 
The three contributions and the total signal calculated for the Hz$\cdots$H\textsubscript{2}O complex are shown in Figure~\ref{fig:hz_h2o_integral}.  
There is an intense SE signal which is centered at 4.40~eV at zero delay time and shifts to lower probe energies within about 20~fs. 
The red-shift reflects the movement of the wave packet in the S\textsubscript{5}/S\textsubscript{6} state from the Franck-Condon zone towards the minimum of the PE surface of the S\textsubscript{5}/S\textsubscript{6} state. 
The signal rapidly becomes weaker in intensity which reflects the decay of the population of the bright S\textsubscript{5}/S\textsubscript{6} state to the dark \textsuperscript{1}$n\pi^{*}$ states S\textsubscript{2}, S\textsubscript{3}, S\textsubscript{4} within about 30~fs. 
Beyond 30~fs, the SE signal continues as a weak tail centered at about 3.50~eV. 
A weak signal also appears at a probe energy of about 2.0~eV and increases in intensity until the end of the simulation, see Figure~\ref{fig:hz_h2o_integral}. 
The decreasing signal at the probe energy of 3.50~eV and the slowly increasing signal at 2.0~eV reflect the decay of the S\textsubscript{5}/S\textsubscript{6} state to the S\textsubscript{1} state. 
While the S\textsubscript{1} state nominally is dark in emission to the S\textsubscript{0} state, it borrows intensity from the bright S\textsubscript{5}/S\textsubscript{6} states. 
Therefore a weak SE signal can be seen from the S\textsubscript{1} state.
The intensity borrowing of the $^{1}n\pi^{*}$ states S\textsubscript{2}, S\textsubscript{3}, S\textsubscript{4} apparently is much weaker.
Therefore such states remain invisible in the SE signal.

The SE signal for the free Hz chromophore is shown in Figure~S4a. 
The intense signal of the bright S\textsubscript{5}/S\textsubscript{6} state is essentially identical with the corresponding signal of the Hz$\cdots$H\textsubscript{2}O complex. 
The weak signal assigned to the S\textsubscript{1} state decays somewhat faster  in free Hz than in the Hz$\cdots$H\textsubscript{2}O complex. 
Presumably, the stabilization of the CT states of \textsuperscript{1}$n\pi^{*}$ and \textsuperscript{1}$\pi\pi^{*}$ character (see Table~S1) by the transfer of the proton from H\textsubscript{2}O to Hz provides additional oscillator strength in the Hz$\cdots$H\textsubscript{2}O complex. 
Remarkably, the SE signal around 2.0~eV probe energy exhibits pronounced oscillations in the probe energy in free Hz, indicating large-amplitude vibrational motion in the hot S\textsubscript{1} state of isolated Hz, see Figure~S4a. 
The amplitude of these oscillations is strongly damped in the Hz$\cdots$H\textsubscript{2}O complex (see Figure~\ref{fig:hz_h2o_integral}a), which is an indication that the hydrogen bond of Hz with H\textsubscript{2}O indeed has a significant effect on the ultrafast radiationless relaxation dynamics in the Hz chromophore that cannot be observed in the population probabilities (Figure~\ref{fig:hz_h2o_adiabatic_pop}).    

The ESA contribution to the integral TA PP signal of the Hz$\cdots$H\textsubscript{2}O complex is depicted in Figure~\ref{fig:hz_h2o_integral}b. 
For the first 20~fs, one observes two intense and rather narrow absorption peaks at $\hbar\omega_{t}$ = 2.2~eV and 2.9~eV. 
Guided by Table~S3, these narrow peaks can be assigned to absorption from the S\textsubscript{5}/S\textsubscript{6} state. 
The redistribution of the intensity from the 2.9~eV peak to the 2.2~eV peak results in an apparent red-shift of the signal. 
Beyond 30~fs, the ESA signal becomes weaker and broader and exhibits irregular fluctuations in the probe energy. 
Overall, the broad signal shifts slightly to the blue, which may reflect the radiationless decay of the $^{1}n\pi^{*}$ states to the S\textsubscript{1} ($\pi\pi^{*}$) state.


Apart from the strong features of the ESA signal, a second, weaker feature can be observed at $\hbar\omega_{\mathrm{pu}} \approx$ 4.5~eV.
This signal is characteristic for transitions from the S\textsubscript{1} ($\pi\pi^{*}$) and S\textsubscript{2} ($n\pi^{*}$) states to higher-lying states.
The weak signal exhibits an oscillation in intensity with a period of about 30~fs with recurrences at \textit{T} = 40, 70 and 100~fs.

The ESA signal for the isolated Hz molecule is shown in Figure~S4b of the SI. 
The signal is overall similar to the signal of the Hz$\cdots$H\textsubscript{2}O complex in Figure~\ref{fig:hz_h2o_integral}b, with the same key features. 
However, the intense signal at small delay times (\textit{T}$\leq$ 25~fs) is weaker in the isolated Hz molecule and subject to faster decay.
The transition energies and oscillator strengths of the ESA transitions are listed in Tables~S2 and S3 for Hz$\cdots$H\textsubscript{2}O and Hz, respectively. 
These data do not provide, however, an obvious explanation of the weaker ESA signal in free Hz.

The GSB signal of the Hz$\cdots$H\textsubscript{2}O complex is displayed in Figure~\ref{fig:hz_h2o_integral}c. 
The position and width of the signal mirror the ground-state absorption spectrum. 
The beatings of the peak intensity reflect ground-state vibrational motion. 
The GSB signal of free Hz shown in Figure~S4c is essentially identical. 
The GSB signal carries little information on the excited-state dynamics and therefore is of minor interest in the present context.

 The total integral TA PP signals of Hz$\cdots$H\textsubscript{2}O and free Hz are shown in Figure~\ref{fig:hz_h2o_integral}d and Figure~S4d, respectively. 
 In both systems, the GSB signal overlaps with the SE signal, but not with the ESA signal, which peaks at lower probe energies. 
 The GSB signal is the strongest of the three contributions and dominates the total signal. 
 For clarity, the GSB contribution has been scaled down by a factor of 10 in Figure~\ref{fig:hz_h2o_integral}d and in Figure~S4d. 
 When comparing the total signals for Hz$\dots$H\textsubscript{2}O and free Hz, one can notice the differences in the ESA signals discussed above. 
 The differences in the SE signals, on the other hand, are overshadowed by the respective GSB signals.

\subsection{2D Electronic Spectra}


The 2D electronic spectra were evaluated with transform limited pump and probe pulses with a duration of 0.1~fs. 
This choice of the pulse duration is motivated by the desire to have a frequency window which covers the complete emission spectrum of the Hz chromophore that extends over several electron volts. 
When using transform-limited pulses, such a short pulse duration is necessary. 
Alternatively, probing with a white-light continuum would allow a broad coverage of the emission spectrum of Hz.

\begin{figure}
    \centering
    \includegraphics[width=0.85\textwidth]{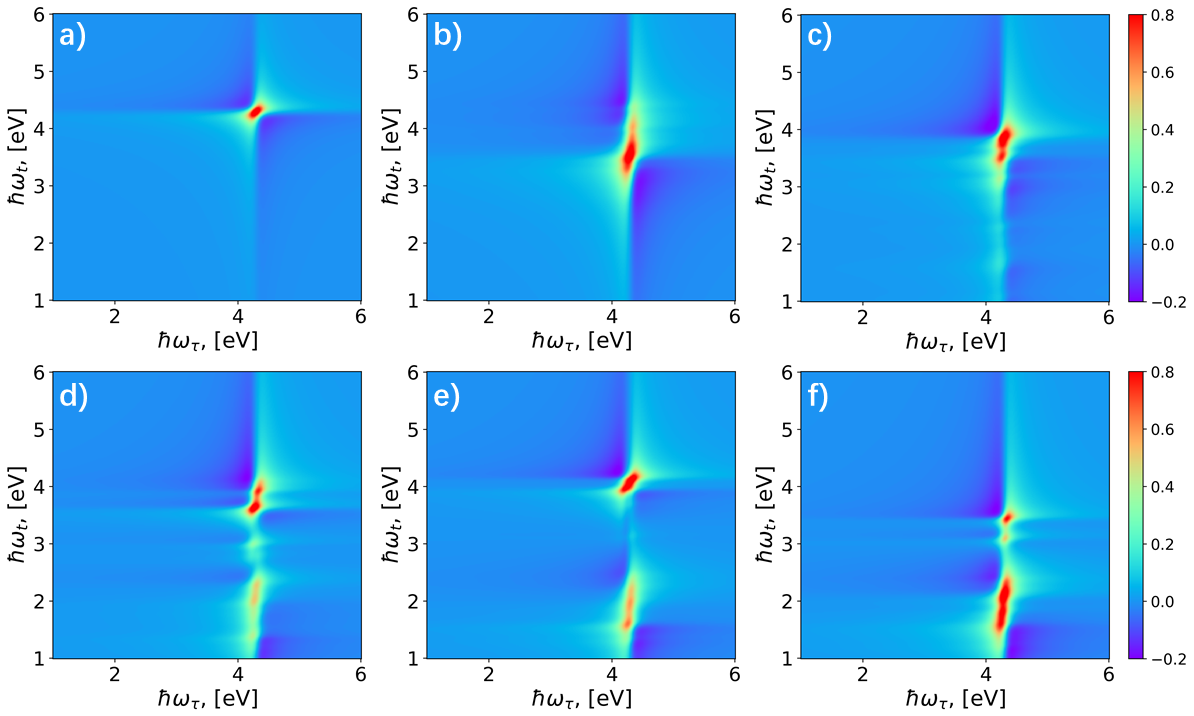}
    \caption{SE contribution to the 2D electronic signal in the rephasing phase-matching direction of the Hz$\cdots$H\textsubscript{2}O complex as a function of the excitation energy $\hbar\omega_{\tau}$ and the detection energy $\hbar\omega_{t}$  at population times $T=0$ (a), 10 (b), 40 (c), 70 (d), 90 (e) and 100~fs (f). The pulse widths are $\tau_{\mathrm{pu}}$ = $\tau_{\mathrm{pr}}$ = 0.1~fs. For each $T$, spectra are rescaled to the same maximum intensity for better visibility.}
    \label{fig:hz_h2o_se_rephasing}
\end{figure}

Figure~\ref{fig:hz_h2o_se_rephasing} shows the 2D SE signals as functions of $\hbar\omega_{t}$ and $\hbar\omega_{\tau}$ for six population times \textit{T} specified in the caption. 
The signals exhibit peaks on the diagonal (Figure~\ref{fig:hz_h2o_se_rephasing}a) or peaks away from the diagonal, which are called cross peaks (Figure~\ref{fig:hz_h2o_se_rephasing}b-f). 
Briefly, diagonal peaks arise when the response originates from the state to which the system was excited, while cross peaks indicate that the response occurs from a state of the system which is different from the state that was excited. 
For better visibility, the signals are rescaled to the same maximum intensity for different waiting times \textit{T}. 
They therefore do not inform on the decay dynamics as function of \textit{T}.

The diagonal peak in Figure~\ref{fig:hz_h2o_se_rephasing}a represents the population of the initially excited bright S\textsubscript{5}/S\textsubscript{6} state. 
The cross peaks in Figures~\ref{fig:hz_h2o_se_rephasing}b,c reflect the red-shift of the SE signal as discussed in the context of Figure~\ref{fig:hz_h2o_integral}a. 
At \textit{T} = 70~fs and later waiting times (Figures~\ref{fig:hz_h2o_se_rephasing}d-f), the stimulated emission from the S\textsubscript{1} state centered at $\hbar\omega_t$ = 2.0~eV becomes visible. 
The large width in $\hbar\omega_t$ reflects the large vibrational excess energy and a large spreading of the vibrational wavepacket in the S\textsubscript{1} state after the internal conversion from the higher electronic states. 
In general, monitoring of the $T$-evolutions of the (relative) intensities of the peaks along $\omega_t$ yields a real-time picture of the population redistribution among the lower-lying excited electronic states. 
The fact that all peaks are located at the same excitation frequency of about 4.4~eV is a peculiar feature of the Hz chromophore. 
Due to the high symmetry of Hz, all excited states in the spectral window of the pump pulse except the S\textsubscript{5}/S\textsubscript{6} state are dark in absorption. 
On the other hand, the ultrafast excited-state dynamics of Hz lowers the symmetry and thus weakens the stringent optical selection rules. 
Therefore, more peaks are seen in emission than in absorption.
Such a detailed picture complements the substantially less detailed view of the process delivered by the SE contribution to the integral TA PP signal (cf. Figure~\ref{fig:hz_h2o_integral}a)).  

The corresponding SE 2D signals of the isolated Hz molecule are shown in Figure~S5a-f. 
The overall picture of the evolution of the SE spectra is similar in Hz and Hz$\cdots$H\textsubscript{2}O.
There is a significant difference though: depopulation of the S\textsubscript{5}/S\textsubscript{6} states and the ensuing population of the S\textsubscript{2} state is much slower in Hz$\cdots$H\textsubscript{2}O than in isolated Hz (cf. panels (d) and (e) of Figures~\ref{fig:hz_h2o_se_rephasing} and S5).
This observation is also corroborated by the comparison of TA PP signals (Figures~\ref{fig:hz_h2o_integral}a and S4a), and can be considered as a signature of the contribution of the hydrogen-bonded water molecule to the excited-state dynamics.
Notably, the signal of the S\textsubscript{5}/S\textsubscript{6} states exhibits a higher intensity than the signal of the S\textsubscript{1}/S\textsubscript{2} states until \textit{T} = 100~fs.
This phenomenon can be explained with the window function $W_{I}$ given in Eq. S11 for the SE contribution in 2D ES.
The large transition dipole moments of the S\textsubscript{5}/S\textsubscript{6} states to the ground state outweighs the larger number of trajectories in the $n\pi^{*}$ and $\pi\pi^{*}$ states.
Note that the SE contribution to the 2D ES signal can also be detected by employing the recently suggested 2D fluorescence-excitation (FLEX) spectroscopy.\cite{Juergen23,pios_2dflex_2024} 
In this signal, the GSB contribution, which overshadows the SE contribution in 2D ES, is absent.


\begin{figure}
    \centering
    \includegraphics[width=0.85\textwidth]{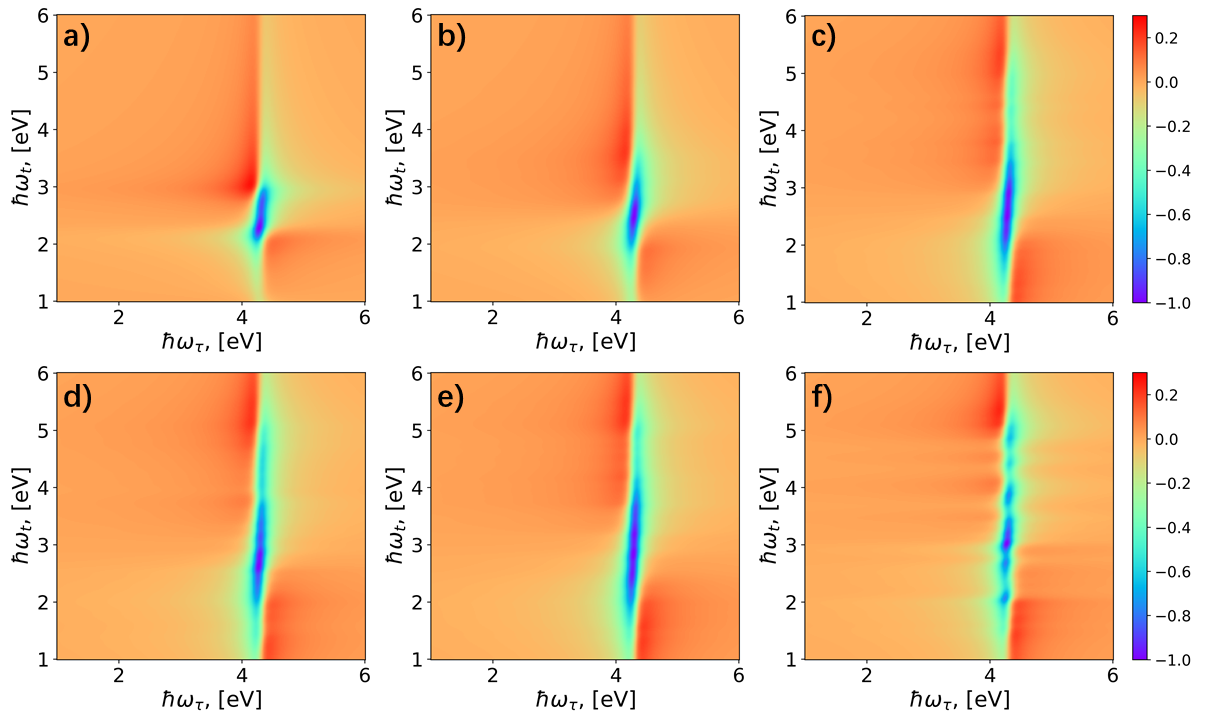}
    \caption{ESA contribution to the 2D electronic signal in the rephasing phase-matching direction of the Hz$\cdots$H\textsubscript{2}O complex as a function of the excitation energy $\hbar\omega_{\tau}$ and the detection energy $\hbar\omega_{t}$  at population times $T=0$ (a), 10 (b), 40 (c), 70 (d), 90 (e) and 100~fs (f). The pulse widths are $\tau_{\mathrm{pu}}$ = $\tau_{\mathrm{pr}}$ = 0.1~fs. For each $T$, the intensities of the spectra are scaled to equal maximum intensities for better visibility.}
    \label{fig:hz_h2o_esa_rephasing}
\end{figure}

The ESA contribution to the 2D electronic spectrum of the Hz$\cdots$H\textsubscript{2}O complex is displayed in Figure~\ref{fig:hz_h2o_esa_rephasing}a-f. 
The ESA signals at the six waiting times are narrowly peaked in $\hbar\omega_{\tau}$ at 4.3~eV. 
In $\hbar\omega_t$, the peaks are much broader and their centers move from $\hbar\omega_t$ = 2.2~eV at \textit{T} = 0~fs towards $\hbar\omega_t$ = 3.0~eV for \textit{T} = 100~fs. 
The broadening of the peaks in $\hbar\omega_t$ reflects the redistribution of the initial electronic-vibrational population in the S\textsubscript{5}/S\textsubscript{6} state over the lower-lying electronic states. 
The interpretation of the shift of the centroid in $\hbar\omega_t$ to higher values than at \textit{T} = 0~fs is difficult, since it depends on the distribution of the oscillator strengths from the states of manifold \{I\} to the states of manifold \{II\}.

\begin{figure}
    \centering
    \includegraphics[width=0.85\textwidth]{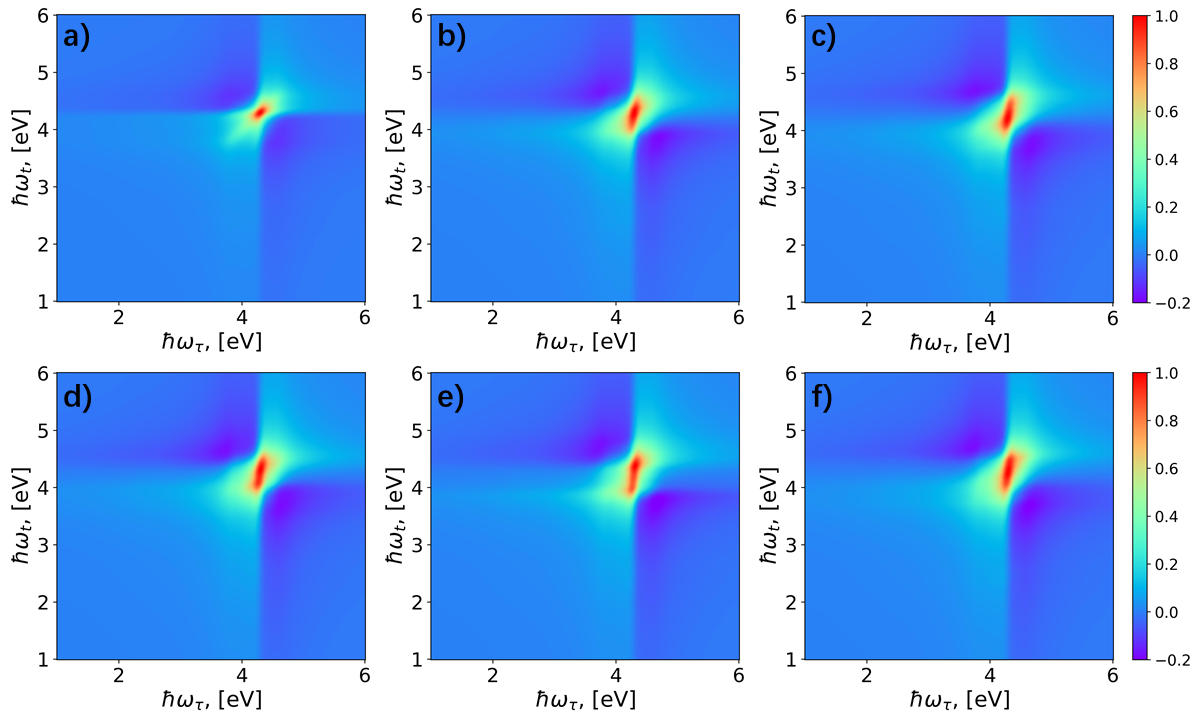}
    \caption{GSB contribution to 2D electronic spectra $I_{\mathrm{2D}}(\omega_{\tau},T,\omega_{t})$ of the Hz$\cdots$H\textsubscript{2}O complex as a function of the excitation energy $\hbar\omega_{\tau}$ and the detection frequency $\hbar\omega_{t}$  at population times $T=0$ (a), 10 (b), 40 (c), 70 (d), 90 (e) and 100~fs (f). For each $T$, the intensities of the spectra are scaled for better visibility.}
    \label{fig:hz_h2o_gsb_rephasing}
\end{figure}

The GSB contribution to the 2D electronic spectrum of the Hz$\cdots$H\textsubscript{2}O complex is shown in Figure~\ref{fig:hz_h2o_gsb_rephasing}a-f.
Like in the case for the TA PP spectrum (Figure~\ref{fig:hz_h2o_integral}c), it is essentially stationary at $\hbar\omega_{\tau} = \hbar\omega_{t}$ = 4.29~eV.
These signals only carry little information on the excited-state dynamics.

\begin{figure}
    \centering
    \includegraphics[width=0.85\textwidth]{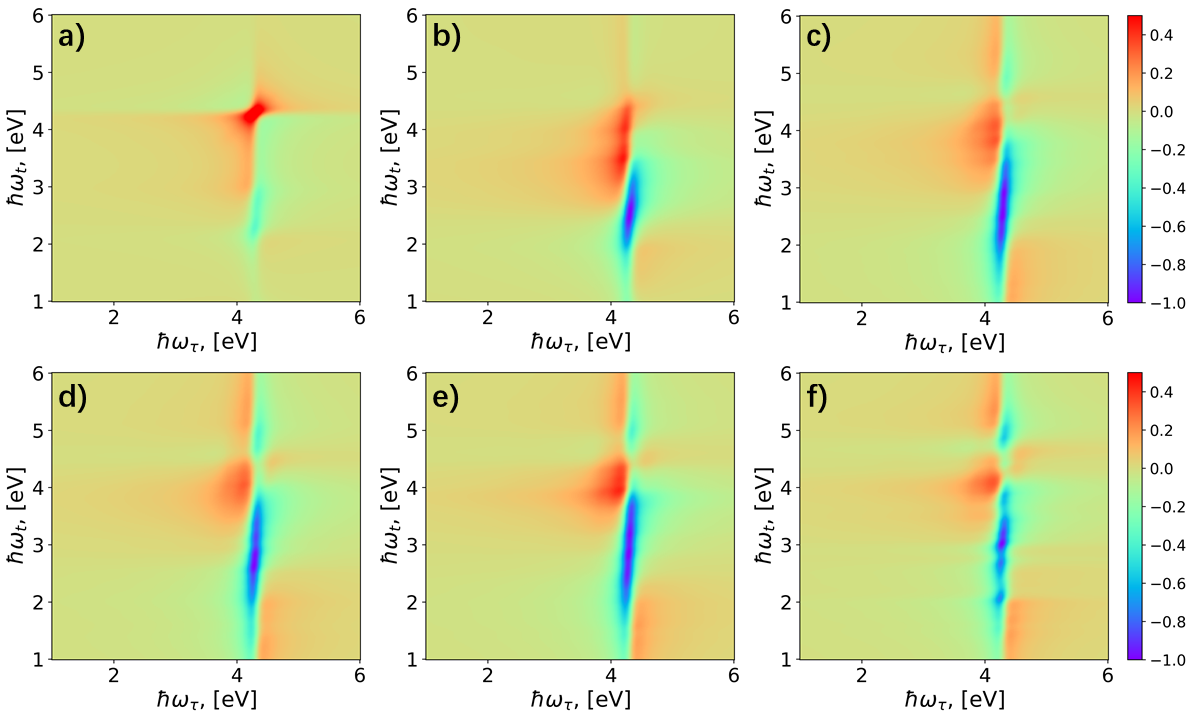}
    \caption{Rephasing 2D electronic spectra $I_{\mathrm{2D}}(\omega_{\tau},T,\omega_{t})$ of the Hz$\cdots$H\textsubscript{2}O complex as a function of the excitation energy $\hbar\omega_{\tau}$ and the detection energy $\hbar\omega_{t}$  at population times $T=0$ (a), 10 (b), 40 (c), 70 (d), 90 (e) and 100~fs (f). For each $T$, the intensities of the spectra are scaled for better visibility. The GSB contribution to the total signal has been reduced by a factor of 10 to improve readability.}
    \label{fig:hz_h2o_sum_rephasing}
\end{figure}

The total 2D electronic spectrum of the Hz$\cdots$H\textsubscript{2}O complex is presented in Figure~\ref{fig:hz_h2o_sum_rephasing}a-f.
For \textit{T} = 10~fs and larger waiting times (Figures~\ref{fig:hz_h2o_sum_rephasing}b-f) a partial cancellation of the GSB and SE signals by the ESA signal occurs, resulting in a signal that can be divided into three parts along the detection energy $\hbar\omega_{t}$.
The first part is a signature of the ESA contribution (negative sign) and extends from $\hbar\omega_{t}$ = 1.5 to 4.0~eV.
The second part is centered around $\hbar\omega_{t} \approx$ 4.3~eV and can be attributed to the GSB contribution (positive sign).
The third  part is detected at $\hbar\omega_{t} \geq$ 4.5~eV and is caused by the upper part of the ESA signal (negative sign).

The SE, ESA, GSB and the total 2D spectra for the isolated Hz molecule are shown in Figures~S5-S8, respectively. 

\section{DISCUSSION AND OUTLOOK}

A mechanistic picture of the photochemistry of the hydrogen-bonded Hz$\cdots$H\textsubscript{2}O has been developed in Ref. \cite{xiang_heptazine2021} and confirmed in the present work with trajectory-based nonadiabatic dynamics simulations. 
The quasi-degenerate S\textsubscript{5}/S\textsubscript{6} state of $\pi\pi^{*}$ character (corresponding to the degenerate S\textsubscript{4} state of free Hz) is the only absorbing state below 5.0~eV. 
The low-lying symmetry-forbidden S\textsubscript{1}($\pi\pi^{*}$) state borrows intensity from the bright S\textsubscript{5}/S\textsubscript{6} state by vibronic coupling and appears as a very weak absorption band at an excitation energy of 2.6~eV (Figure~\ref{fig:hz_h2o_uvvis}). 
Excitation of the Hz$\cdots$H\textsubscript{2}O complex by a short UV pulse prepares a wave packet in the S\textsubscript{5}/S\textsubscript{6} state. 
The population of the S\textsubscript{5}/S\textsubscript{6} states decays within about 20~fs to the states S\textsubscript{2}, S\textsubscript{3}, S\textsubscript{4} of $n\pi^{*}$ character. 
The population transfer within 20~fs is approximately sequential, that is, S\textsubscript{4} is populated first, then S\textsubscript{3} and finally S\textsubscript{2}. 
The population of the S\textsubscript{1} state by radiationless relaxation begins at about 20~fs and then increases monotonically, reaching 80\% at 100~fs (Figure~\ref{fig:hz_h2o_adiabatic_pop}). 
The decay of the S\textsubscript{1} state to the electronic ground state is very slow, resulting in a population probability of the S\textsubscript{0} state of just a few per cent at 100~fs. 
The vibrationally hot S\textsubscript{1} state thus acts a reservoir in which a significant part of the energy of the absorbed photon is stored for at least tens or hundreds of picoseconds due to very low fluorescence and internal conversion rates, as observed experimentally for a derivative of Hz, trianisoleheptazine (TAHz).\cite{rabe_tahz_2018} 
The experimentally observed liberation of OH radicals indicates that water molecules can be oxidized by a PCET reaction in the long-lived S\textsubscript{1} state of the TAHz$\cdots$H\textsubscript{2}O complex.\cite{rabe_tahz_2018} 

In the present work, we wanted to explore to what extent the complex photochemistry of the Hz$\cdots$H\textsubscript{2}O complex on femtosecond timescales can be visualized by TA PP and 2D electronic spectroscopy. 
While the interpretations of the individual signals were discussed above in the Results section, we summarize these findings here by focusing on the individual electronic states of the Hz$\cdots$H\textsubscript{2}O complex.

As to be expected, the population of the bright S\textsubscript{5}/S\textsubscript{6} state shows up as an intense signal in the SE contribution to the TA PP spectrum (Figure~\ref{fig:hz_h2o_integral}a). 
The time evolution of the intensity of the signal monitors the decay of the S\textsubscript{5}/S\textsubscript{6} population within about 20~fs, while the red-shift of the signal in $\hbar\omega_{t}$ reports the movement of the photoexcited wave packet towards the equilibrium geometry of the S\textsubscript{5}/S\textsubscript{6} state. 
The weak oscillatory signal extending beyond $\approx$~40~fs reflects remnants of the wave packet remaining in the S\textsubscript{5}/S\textsubscript{6} state. 
The population of the S\textsubscript{5}/S\textsubscript{6} state also shows up prominently at short delay times in the ESA component of the TA PP signal (Figure~\ref{fig:hz_h2o_integral}b). 
According to the electronic-structure calculations (Table~S3), there exist two optical transitions from the S\textsubscript{5}/S\textsubscript{6} state with significant oscillator strengths near 2.2~eV and 2.9~eV. 
These are seen as strong absorptive signals in Figure~\ref{fig:hz_h2o_integral}b existing up to about 20~fs. 
The weak recurring signal at $\hbar\omega{t} \approx$ 4.6~eV at \textit{T} = 40, 70 and 100~fs can in part be explained by the revival of the adiabatic S\textsubscript{2} state (Figure~\ref{fig:hz_h2o_adiabatic_pop}). 
This conclusion is based on a bright transition from the S\textsubscript{2} to the S\textsubscript{45} state at 4.57~eV at the ground-state equilibrium geometry, which is assumed to come down, entering the manifold of computed states.
The 2D SE signal also monitors the dynamic redshift of the excitation energy of the S\textsubscript{5}/S\textsubscript{6} state (Figure~\ref{fig:hz_h2o_se_rephasing}a-c). 
Beyond about 30~fs, the peak position in $\hbar\omega_{t}$ exhibits vibrational oscillations in qualitative agreement with Figure~\ref{fig:hz_h2o_integral}a (it should be recalled that the intensities of the 2D spectra are rescaled and therefore do not reflect the evolution of the intensities with respect to the waiting time). 
In the 2D ESA signal (Figure~\ref{fig:hz_h2o_esa_rephasing}), the population of the S\textsubscript{5}/S\textsubscript{6} state is monitored by a peak at $\hbar\omega{t} \approx$ 2.2~eV which rapidly broadens and shifts to higher energy with increasing waiting time. 
Due to its large width in $\hbar\omega_{t}$, this signal is rather unspecific and may include ESA transitions from other intermediately populated electronic states. 
The population dynamics and the vibrational dynamics of the initially populated S\textsubscript{5}/S\textsubscript{6} state is thus authentically reported by several of the time and energy resolved signals considered herein.

The detection of the population dynamics and the vibrational dynamics of the intermediately populated $^{1}n\pi$ states is more challenging. 
In the SE component of the TA PP signal (Figure~\ref{fig:hz_h2o_integral}a), the region in $\hbar\omega_{t}$ in which the SE from the $^{1}n\pi$ states is expected is notably dark. 
Apparently, the vibronic intensity borrowing of the $^{1}n\pi^{*}$ states from the bright S\textsubscript{5}/S\textsubscript{6} state is too weak to yield a detectable signal. 
From Table~S3, an ESA signal of the $^{1}n\pi^{*}$ states is expected near 2.0~eV with significant intensity. 
The signal in Figure~\ref{fig:hz_h2o_integral}b extending from about 25~fs to about 50~fs and fading away afterwards could arise from ESA from the $^{1}n\pi^{*}$ states, but it could also arise from remnants of the population of the S\textsubscript{5}/S\textsubscript{6} state. 
The 2D SE signals likewise do not show clear evidence of the transient population of $^{1}n\pi^{*}$ states. 
The 2D ESA signals are too broad and structureless in $\hbar\omega_{t}$ to provide clear evidence of the $^{1}n\pi^{*}$ states.

More important than the monitoring of the transiently populated $^{1}n\pi^{*}$ states is the monitoring of the population of the long-lived S\textsubscript{1} ($\pi\pi^{*}$) reservoir state. 
In the SE component of the TA PP signal (Figure~\ref{fig:hz_h2o_integral}a), weak emission from the S\textsubscript{1} state is clearly observed near 2.0~eV in $\hbar\omega_{t}$. 
This signal becomes visible at \textit{T} = 30~fs and steadily increases in intensity with the delay time, mirroring the monotonous increase of the population of the S\textsubscript{1} state (Figure~\ref{fig:hz_h2o_adiabatic_pop}). 
The signal is relatively weak due to the optically forbidden character of the S\textsubscript{1}-S\textsubscript{0} transition. 
It is interesting to see that the vibronic intensity borrowing in Hz is sufficiently strong to generate a detectable PP SE signal of the dark S\textsubscript{1} state. 
The PP ESA signal of the S\textsubscript{1} state is expected near 3.5~eV in $\hbar\omega_{t}$ with significant oscillator strength (Table~S3). 
There is no clear signature of this expected contribution in Figure~\ref{fig:hz_h2o_integral}b. 
In the 2D SE spectrum, the population of the S\textsubscript{1} state is reflected by the signal near 2.0~eV in $\hbar\omega_{t}$, see Figure~\ref{fig:hz_h2o_se_rephasing}d-f. 
The large width of the signal in $\hbar\omega_{t}$ reflects the large vibrational excess energy in the S\textsubscript{1} state. 
In the 2D ESA spectrum, a strong signal at the expected transition energy of 3.5~eV is observed at 40~fs and beyond, but the very large width of the ESA signals in Figure~\ref{fig:hz_h2o_esa_rephasing}d-f does not allow assignments to specific electronic states.

The PCET reaction on the PE surface of the S\textsubscript{1} state in the Hz$\cdots$H\textsubscript{2}O complex is hindered by a barrier which has been estimated as 0.98~eV.\cite{ehrmaier_jpca_2020_molecular}
The excited-state PCET reaction thus is in the deep tunneling regime and the expected tunneling time is far too long to be explored with \textit{ab initio} on-the-fly dynamics. 
However, when the vibrationally hot population of the S\textsubscript{1} state is re-excited by an additional laser pulse (a so-called “push” pulse) to a higher electronic state, the PCET reaction becomes ultrafast and can be probed in a femtosecond “pump-push-probe” experiment. 
The push pulse speeds up the PCET reaction to an extent that the dynamics and the time-resolved spectroscopy of the pump-push-probe experiment can be simulated with \textit{ab initio} nonadiabatic molecular dynamics calculations. 
A proof-of-principle demonstration of a pump-push-probe experiment was reported by Corp et al.\cite{corp_jpcc_2020} for complexes of TAHz with phenol molecules using picosecond laser pulses. 
In a future continuation of the present work, pump-push-probe experiments with femtosecond laser pulses will be simulated for the Hz$\cdots$H\textsubscript{2}O complex.

\section{Supporting Information}

Methodology, results for isolated Hz, additional vertical excitation energies, ESA contributions at the ground-state equilibrium geometry, doorway-window approximation

\begin{acknowledgement}
S.V.P. and L.P.C. acknowledge support from the National Natural Science Foundation of China (No.~22473101). 
M. F. G. acknowledges support from the National Natural Science Foundation of China (No.~22373028). 
\end{acknowledgement}
\section{Data availability statement}
The data that supports the findings of this study is available from the corresponding author upon reasonable request.
\section{Code availability statement}
The code to generate the transient-absorption and 2D electronic spectra is available under \url{https://github.com/psebastianzjl/pumpprobe_spectroscopy}.
\section{Competing interests}
The authors declare that they have no competing interests.
%
%
\bibliography{main}
\end{document}